\documentclass[12pt]{iopart}
\input epsf

\usepackage[dvips]{color}

\usepackage{amsfonts}
\usepackage{amssymb}
\usepackage{amsbsy}

\newcounter{append}
\newcommand{\bc}{\begin{center}}
\newcommand{\ec}{\end{center}}
\newcommand{\be}{\begin{equation}}
\newcommand{\ee}{\end{equation}}
\newcommand{\ba}{\begin{array}}
\newcommand{\ea}{\end{array}}
\newcommand{\beqn}{\begin{eqnarray}}
\newcommand{\eeqn}{\end{eqnarray}}

\def\moyenne#1{\langle #1 \rangle}

\begin{document}
\jl{1}

\title{Relaxation in the XX quantum chain}
\author{Thierry Platini and Dragi Karevski
\footnote{Corresponding author: {\tt karevski@lpm.u-nancy.fr}}
 }
\address{Laboratoire de Physique des Mat\'eriaux, 
	UMR CNRS 7556,\\ 
  Universit\'e Henri Poincar\'e, Nancy
  1,\\ B.P. 239,
  F-54506  Vand\oe uvre les Nancy Cedex, France
}
\date{\tt \today}

 %%%%%%%%%%%%%%%%%%%%%%%%%%%%%%%%%%%%%%%%%%%%%%%%%%%%%%%%%%%%%%%%%%%%%%%%%%%%%%%%
\begin{abstract}
We present the results obtained on the magnetisation relaxation properties of an XX quantum chain in a transverse magnetic field. We first consider an initial thermal kink-like state where half of the chain is initially thermalized at a very high temperature $T_b$ while the remaining half, called the system, is put at a lower temperature $T_s$. From this initial state, we derive analytically the Green function associated to the dynamical behaviour of the transverse magnetisation. Depending on the strength of the magnetic field and on the temperature of the system, different regimes are obtained for the magnetic relaxation. In particular, with an initial droplet-like state, that is a cold subsystem of finite size  in contact at both ends with an infinite temperature environnement, we derive analytically the behaviour of the time-dependent system magnetisation.
\end{abstract}
%%%%%%%%%%%%%%%%%%%%%%%%%%%%%%%%%%%%%%%%%%%%%%%%%%%%%%%%%%%%%%%%%%%%%%%%%%%%%%%%

%%%%%%%%%%%%%%%%%%%%%%%%%%%%%%%%%%%%%%%%%%%%%%%%%%%%%%%%%%%%%%%%%%%%%%%%%%%%%%%%
\section{Introduction}
%%%%%%%%%%%%%%%%%%%%%%%%%%%%%%%%%%%%%%%%%%%%%%%%%%%%%%%%%%%%%%%%%%%%%%%%%%%%%%%%
Non-equilibrium properties of quantum chains generated by initial inhomogeneities, or by the sudden variation of an external field have been the purpose of several extended studies starting from the sixties and continuing up to nowadays. In particular, in the context of free-fermionic models, that is Ising-like chains \cite{lieb,karebook}, since the dynamics can be solved exactly \cite{jacoby1,jacoby2}, a lot of exact results have been obtained. One may mention  magnetisation relaxation profiles which show a  scaling behaviour \cite{berim1,berim2,berim,berim3,cabrera,schutztrimper,igloirieger,karevski1,ogat,karebook},  two-point correlation functions characterising ageing properties \cite{schutztrimper,igloirieger},  magnetisation and energy current densities \cite{antal1,antal2,antal3,popkov,eisler}, and one may notice that a quantized relaxation behaviour was also observed\cite{hunyadi,platini1,platini2}. 
Since the very first studies of Niemejer and Tjion and others \cite{niemeijer,tjion,barouch1,barouch2}, the relaxation obtained shows typically a very slow decay. Indeed, they have shown that the relaxation of the magnetisation in a spin-free fermionic chain after a sudden change of the external magnetic field is algebraic instead of exponential \cite{terwielmazur} and that there is a lack of ergodicity in the sense that the magnetisation does not relaxe toward its equilibrium value \cite{barouch1}. Consequently, in such systems there is no finite relaxation time. Moreover, it was shown that for certain inhomogeneously magnetised initial states, at special wave vectors, a slowing down of the magnetisation relaxation can occur for gapped free fermionic models \cite{cabrera}. The effect of quenched disorder was also considered in the Ising quantum chain \cite{abriet1}. Once more, for an initially completely ordered state, either in the field direction or perpendicular to it, an algebraic decay of the magnetisation toward a stationary state was obtained. 
More recently, in the critical Ising quantum chain we have studied the relaxation of an initially magnetised finite droplet at a given temperature in contact at both ends with infinite Ising chains. We have shown that the droplet magnetisation decays in time to the equilibrium magnetisation with a $1/tf(x/t)$ behaviour \cite{platini1} where $f(u)$ is a scaling function depending on the temperature. For a vanishing droplet initial temperature, the scaling function is simply given by the characteristic function $\Pi(u)$. 

In the present study, we focus our attention to the same situation in the $XX$ quantum chain in a transverse magnetic field. Whereas in the Ising case, the transverse magnetisation is not conserved in time, in the $XX$ quantum chain it is a conserved dynamical quantity and one may expect that this conservation law will show up in a way or an other in the relaxation behaviour. 
To be more specific, let us precise the initial set-up. The system is initially prepared in a factorized canonical state $\rho(0)=\rho_b(0)\rho_s(0)$ where the density matrices 
$\rho_{b,s}(0)\propto e^{-\beta_{b,s}{\cal H}_{b,s}}$ are canonical states of parts of the (infinite) chain each thermalized at a given temperature $\beta_{b,s}^{-1}$. At time $t=0$, the (local) interactions between the different parts are switched on and the dynamics is generated by the total Hamiltonian ${\cal H}={\cal H}_b+{\cal H}_s+{\cal H}_I$. 
The paper is organised as follows: In the next section we give some details of the by now classical diagonalisation of the $XX$-model and its dynamics.  Section~3 gives more details on the initial set-up and on the time-dependent expectation values of physical quantities, putting a special focus on the transverse magnetisation. In particular, it is shown that the time-dependent transverse magnetisation may be expressed as a discret convolution product of a Green function with the initial magnetisation profile. In section~4, from an initial kink-like state we derive analytically the magnetisation Green function and show that, at vanishing system temperature, one has to distinguish between the low and high field regimes. Consequently, in section~5, we show that the relaxation of the magnetisation of an initial finite droplet of size $L_s$, toward its equilibrium vanishing value, changes after an initial linear decay from an algebraic decay, $L_s/t$, at large fields to a faster size-independent decay at lower fields. However, this behaviour does not survive for finite system temperatures, but nevertheless the magnetisation relaxation at low enough temperatures is strongly affected. 
We finally summarize our results and draw out some conclusions.

%%%%%%%%%%%%%%%%%%%%%%%%%%%%%%%%%%%%%%%%%%%%%%%%%%%%%%%%%%%%%%%%%%%%%%%%%%%%%%%%
\section{Diagonalisation and Dynamics}
%%%%%%%%%%%%%%%%%%%%%%%%%%%%%%%%%%%%%%%%%%%%%%%%%%%%%%%%%%%%%%%%%%%%%%%%%%%%%%%%
\subsection{Diagonalisation}
The free boundary conditions XY quantum chain in a transverse field is defined through the Hamiltonian
	\begin{equation}
		{\cal H}=- \frac {1}{2} \sum_{n=1}^{L-1} \left[ J_x\sigma_n^x \sigma_{n+1}^x+J_y\sigma_n^y \sigma_{n+1}^y\right]
		- \frac {h}{2} \sum_{n=1}^{L} \sigma_{n}^z
		\label{eq211}
	\end{equation}
where the $\sigma$s are the Pauli matrices at site $n$, $h$ is the transverse field applied in the z-direction and where we parametrize the coupling constants using the anisotropy factor $\kappa\in[-1,1]$ as $J_x=(1+\kappa)/2$ and $J_y=(1-\kappa)/2$. In the special case $\kappa=1$, the $y$-direction coupling constants vanish and one recovers the Ising quantum chain case. At $\kappa=0$, $J_x=J_y$ and the Hamiltonian describes the $XX$-model for which the total transverse magnetisation $M^z=\sum_n \sigma^z_n$ is a conserved quantity since we have $[{\cal{H}},M^z]=0$. Consequently,  the exchange interaction energy and the Zeeman energy are separatly conserved quantities. 

In terms of the lattice Clifford's operators $\Gamma_{n}^{1}$, $\Gamma_{n}^{2}$
	\begin{eqnarray}
	\Gamma_{n}^{1} &=& \prod_{j=1}^{n-1} (-\sigma_{j}^{z}) \sigma_{n}^{x}\nonumber\\
	\Gamma_{n}^{2} &=& -\prod_{j=1}^{n-1} (-\sigma_{j}^{z}) \sigma_{n}^{y} 	\label{eq212}
	\end{eqnarray}
with $\Gamma_{n}^{i \dag}=\Gamma_{n}^{i}$ satisfying the anticommutation rules
	\begin{eqnarray}
	\{ \Gamma_{n}^{i} ,\Gamma_{m}^{j} \} &=& 2 \delta_{i,j} \delta_{n,m}
	\label{eq213}
	\end{eqnarray}
the Hamiltonian (\ref{eq211}) takes the quadratic form \cite{jord-wigner,lieb,karebook}
	\begin{equation}
		{\cal H}=- \frac {1}{4} \mathbf{\Gamma^{\dag}} \mathbf{T} \mathbf{\Gamma}
	\label{eq214}
	\end{equation}
where  $\mathbf{\Gamma}$ is the $2L$-component vector
	\begin{eqnarray}
	\mathbf{\Gamma}=\left(\begin{array}{c}
	\Gamma^1 \\  
	\Gamma^2
	\end{array}\right),
\ {\rm with} \ \
	\mathbf{\Gamma^i}=\left(\begin{array}{c}
	\Gamma^i_{1}\\
	\Gamma^i_{2}\\
	\vdots\\ 
	\Gamma^i_{L}
	\end{array}\right),
\  {\rm and} \ i=1,2 \; .
\label{eq215}
	\end{eqnarray}
The $2L\times 2L$ hermitian matrix $\mathbf{T}$ is given by
	\begin{eqnarray} 
	\mathbf{T}=\left(\begin{array}{cc}
	0 & \mathbf{C} \\ 
	\mathbf{C^{\dag}} & 0	 
	\end{array}\right)
	\label{eq216}
	\end{eqnarray}	
with
	\begin{eqnarray} 
	\mathbf{C}=-i
	\left(\begin{array}{ccccc}
	h & J_y &  & &0\\
	J_x & h & J_y &  & \\
	   & \ddots &\ddots &\ddots &\\
	   
	    &  &  J_x &h & J_y\\
	  0  & &    & J_x &h\\
	\end{array}\right).
	\label{eq217}
	\end{eqnarray}
Parametrizing the eigenvectors $V_q$, associated to the eigenvalue $\epsilon_q$, of $\bf T$ as
	\begin{eqnarray} 
	V_q=\frac{1}{\sqrt{2}}
	\left(\begin{array}{c}
	\phi_q \\
	-i\psi_q\\
	\end{array}\right)
	\label{eq218}
	\end{eqnarray}
we obtain from the eigenvalue equation ${\bf T}V_q=\epsilon_q V_q$ the following equations:
\be
-i{\bf C}\psi_q=\epsilon_q\phi_q,\qquad {\bf C}^{\dag}\phi_q=-i\epsilon_q\psi_q\; .
\label{eq219}
\ee
We can notice here that if we change simultaneously $\epsilon_q\rightarrow -\epsilon_q$ and
$\psi_q\rightarrow-\psi_q$ these equations are unchanged. So, to each eigenvalue $\epsilon_q\ge 0$ corresponds an eigenvalue $\epsilon_{q'}=-\epsilon_q$ with the associated eigenvector $V_{q'}(\phi_{q'},\psi_{q'})=V_q(\phi_q,-\psi_q)$.

Introducing the fermionic creation and annihilation operators $\eta_{q}^{\dag}$, $\eta_{q}$ 
	\begin{eqnarray}
\eta_{q}^{\dag}=\frac{1}{2}\sum_n\left(\phi_q(n)\Gamma_{n}^{1}-i\psi_q(n)\Gamma_{n}^{2}\right) \nonumber\\
\eta_{q}=\frac{1}{2}\sum_n\left(\phi_q(n)\Gamma_{n}^{1}+i\psi_q(n)\Gamma_{n}^{2}\right)
	\label{eq2110}
	\end{eqnarray}
satisfying the canonical Fermi-Dirac anticommutation rules $\{\eta_q^{\dag},\eta_{q'}\}=\delta_{q,q'}$, the Hamiltonian (\ref{eq211}) takes the free-fermionic structure	
	\begin{equation}
		{\cal H}=\sum_{q=1}^{L}\epsilon_q\left(\eta_{q}^{\dag}\eta_{q}-1/2\right)
	\label{eq2111}
	\end{equation}
where $\epsilon_q$ are the positive eigenvalue of $\mathbf{T}$, given by
\be
	\epsilon_q=\sqrt{\kappa^2\sin^2(q)+(h+\cos(q))^2}\; .
\label{eq2112}
\ee

In the XX-quantum chain, with $\kappa=0$, one has for the (positive) spectrum
\be
\epsilon_q=|h+\cos q|=|h-\cos k|
\label{eq2113}
\ee
with $k=\pi-q\in[0,\pi]$ and where the gap vanishes for $h\le 1$ at $k_F=\arccos (h)$.
In an equivalent picture, the (positive) spectrum associated to $k\in[0,k_F]$ is usually interpreted as corresponding to the creation of holes in the Fermi sea, while the ground state corresponds to $\prod_{k=0}^{k_F}b^{\dag}_k|0\rangle$ where the $b^{\dag}$ are particle fermionic creation operators with negative energies $\varepsilon_k=-\epsilon_k$. 
	
\subsection{Dynamics}
The Heisenberg equations of motion  for the lattice Clifford's operators are easily solved, see the appendix~1 for more details, and lead to \cite{karebook}
	\begin{eqnarray} 
	\Gamma_n^j(t)\equiv e^{i{\cal H} t}\Gamma_n^j e^{-i{\cal H} t}=\sum_{k,\nu} \langle\Gamma_{k}^{\nu} |\Gamma_{n}^{j}(t)\rangle \Gamma_k^\nu
	\label{Gamma-t}
\label{eq221}
	\end{eqnarray}
where $\langle C |D \rangle=\frac{1}{2}\{C^{\dag},D\}$ is a pseudo-scalar product with $\{.,.\}$ the anticommutator. The time-dependent contractions are explicitely given by
	\begin{eqnarray} 
	\langle\Gamma_{k}^{1} |\Gamma_{n}^{1}(t)\rangle&=&\sum_q\phi_q(k)\phi_q(n)\cos(\epsilon_qt)\nonumber \\
	\langle\Gamma_{k}^{2} |\Gamma_{n}^{2}(t)\rangle&=&\sum_q\psi_q(k)\psi_q(n)\cos(\epsilon_qt)\nonumber \\
	\langle\Gamma_{k}^{1} |\Gamma_{n}^{2}(t)\rangle&=&\langle\Gamma_{n}^{2} |\Gamma_{k}^{1}\rangle_{-t}=-\sum_q\phi_q(k)\psi_q(n)\sin(\epsilon_qt)\; . 
	\label{eq222}
	\end{eqnarray}
In the XX-chain case, in the thermodynamic limit they take the form
\begin{eqnarray}
	\langle \Gamma_k^1|\Gamma_n^1(t)\rangle=i^{n-k}J_{n-k}(t)\left\{
	\begin{array}{cc}
	\cos(ht);&n-k=2p\\
	-i\sin(ht);&n-k=2p+1
	\end{array}\right.\\
	\langle \Gamma_k^1|\Gamma_n^2(t)\rangle=i^{n-k}J_{n-k}(t)\left\{
	\begin{array}{cc}
	-\sin(ht);&n-k=2p\\
	-i\cos(ht);&n-k=2p+1
	\end{array}\right.	
	\label{eq223}
\end{eqnarray}
with $\langle \Gamma_k^2|\Gamma_n^2(t)\rangle =\langle \Gamma_k^1|\Gamma_n^1(t)\rangle$ and where $J_n(t)$ is the J-Bessel function of order $n$. These contractions are the basic time-dependent quantities of the $XX$-chain.

Formally, since $\langle \Gamma_k^i | \Gamma_n^j\rangle=\delta_{i,j}\delta_{n,k}$,
the set of $\{\Gamma_k^i\}$ forms an orthonormal basis of a 2L-dimensional
linear vector space $\cal E$ with inner product defined by $\langle .| .\rangle=\frac{1}{2}\{.^\dag,.\}$. Hence, every vector $X\in {\cal E}$ has a unique expansion $X=\sum_{i,k}\langle \Gamma_k^i|X\rangle \Gamma_k^i$. 
A string operator of the form ${\cal O}\equiv X_1X_2\dots X_n$ is a direct product vector in ${\cal E}^{\otimes n}$ and has the expansion
\begin{equation}
{\cal O}\equiv  X_1X_2\dots X_n=\sum_{i_1,k_1}\dots \sum_{i_n,k_n} \langle \Gamma_{k_1}^{i_1} | X_1\rangle \dots \langle \Gamma_{k_1}^{i_n} | X_n\rangle
\Gamma_{k_1}^{i_1}\dots\Gamma_{k_n}^{i_n}\; .
\label{eq224}
\end{equation}  
%%%%%%%%%%%%%%%%%%%%%%%%%%%%%%%%%%%%%%%%%%%%%%%%%%%%%%%%%%%%%%%%%%%%%%%%%%%%%%%%

%%%%%%%%%%%%%%%%%%%%%%%%%%%%%%%%%%%%%%%%%%%%%%%%%%%%%%%%%%%%%%%%%%%%%%%%%%%%%%%%
\section{Initial conditions and expectation values}
%%%%%%%%%%%%%%%%%%%%%%%%%%%%%%%%%%%%%%%%%%%%%%%%%%%%%%%%%%%%%%%%%%%%%%%%%%%%%%%%
\subsection{Product initial state}
The system is initially prepared in a factorized state (a product measure state) \cite{schutztrimper,karevski1,ogat,platini1,platini2} of the form
	\begin{eqnarray} 
	\rho(0)=\prod_j \rho_j(0)
	\label{eq311}
	\end{eqnarray}
where the density matrix 
	\begin{eqnarray} 
	\rho_j(0)=\frac{1}{Z_j}\exp(-\beta_j{\cal{H}}_j)
	\label{eq312}
	\end{eqnarray}
is a canonical state, at inverse temperature $\beta_j$, associated to the Hamiltonian ${\cal H}_j$ of the $j$th subsystem of the total chain decomposed in the following way:
	\begin{equation}
		{\cal H}=\sum_j {\cal H}_j +\sum_j {\cal H}^I_{j,j+1}\; .
	\label{eq313}
	\end{equation}
where the Pauli's matrices of the $j^{th}$ part are written $(\sigma_n)_j$. The interface interaction term is
	\begin{equation}
	{\cal H}^I_{j,j+1}=-\frac{1}{4}\left[(\sigma_{L_j}^x)_j (\sigma_{1}^x)_{j+1} +(\sigma_{L_j}^y)_j (\sigma_{1}^y)_{j+1}\right]\; ,
	\label{eq314}
	\end{equation}
where we wrote the Pauli matrices of the $j^{th}$ part as $(\sigma_n)_j$. 

\subsection{Expectation value of the transverse magnetisation}
The expectation value of an observable $\cal{O}$ at a time $t$ is given by
	\begin{eqnarray} 
	\moyenne{{\cal{O}}(t)}=Tr\{{\cal{O}}(t)\prod_j \rho_j(0)\}\; ,
	\label{eq321}
	\end{eqnarray}
where ${\cal{O}}(t)=e^{i{\cal H}t}{\cal O}e^{-i{\cal H}t}$ is the operator associated to the observable at time $t$ in the Heisenberg picture.
In particular if $\cal{O}$ is a string of Clifford's operators
${\cal{O}}=\Gamma_{l_1}^{j_1}\dots\Gamma_{l_n}^{j_n} $,  one has at time~$t$
	\begin{eqnarray} 
	\moyenne{{\cal{O}}(t)}= \sum_{i_1,k_1}\dots \sum_{i_n,k_n} \langle \Gamma_{k_1}^{i_1} | \Gamma_{l_1}^{j_1}(t)\rangle \dots \langle \Gamma_{k_1}^{i_n} |\Gamma_{l_n}^{j_n}(t) \rangle
Tr\{\Gamma_{k_1}^{i_1}\dots\Gamma_{k_n}^{i_n}\rho(0)\}\; ,
	\label{eq322}
	\end{eqnarray}
where the elements $Tr\{\Gamma_{k_1}^{i_1}\dots\Gamma_{k_n}^{i_n}\rho(0)\}$ represent the relevant initial state information (correlations) for the time evolution of the quantity $\cal O$. Since the time-contractions are explicitely known one needs only to compute the initial state properties of the string operators for a given $\rho(0)$.
	
In this work we focus our attention on the time behaviour of the transverse magnetisation which is, as already stated, a conserved quantity.
The transverse magnetisation is one of the simplest Clifford string operator. Namely, it takes the form $\sigma_{n}^{z} = -i\Gamma_{n}^{2}\Gamma_{n}^{1}$ and at time $t$, in 
the Heisenberg picture we have
	\begin{equation}
 	\sigma_{n}^{z}(t) = -i\Gamma_{n}^{2}(t)\Gamma_{n}^{1}(t)\; .
	\label{eq323}
	\end{equation}
Its expectation value, in the thermodynamic limit, at time $t$ is given by
	\begin{equation}
	\moyenne{\sigma_{n}^{z}}(t)=\sum_{k,j} F_t^j(n,k) I_{k,k+j}\; 
	\label{Mz=f(I)}
	\label{eq324}
	\end{equation}
where $I_{k,k+j}=Tr\{-i\Gamma_k^2\Gamma_{k+j}^1\rho(0)\}$ is the initial condition matrix. The time-dependent functions $F^j_t$ are given by 
\begin{equation}
	F^{j}_t(n,k)=\langle \Gamma_k^1|\Gamma_n^1(t)\rangle\langle\Gamma_{k+j}^2|\Gamma_n^{2}(t)\rangle
	-\langle\Gamma_k^1|\Gamma_n^2(t)\rangle\langle\Gamma_{k+j}^2|\Gamma_n^{1}(t)\rangle\; .
	\label{eq325}
\end{equation}
and they can be explicitely evaluated in the thermodynamic limit:
\begin{equation}
F^{2j}_t(n,k)=F^{2j}_t(n-k)=(-1)^jJ_{n-k}(t)J_{n-k-2|j|}(t)\; ,
	\label{eq326}
\end{equation}
and they identically vanish for odd $j$. Consequently, we have for the time behaviour of the transverse magnetisation
\begin{equation}
	\moyenne{\sigma_{n}^{z}}(t)=\sum_{j=-\infty}^{\infty} \left(F_t^{2j}\star P^{2j}\right)(n) \; ,
	\label{eq327}
\end{equation}
where $\star$ stands for a discret convolution product and where we introduced the notations $P^{2j}(k)=I_{k,k+2j}$.
 
One needs now to evaluate the initial matrix $I$. First remark that for
Clifford operators $\Gamma$ belonging to different initial blocks, the trace identically vanishes, reflecting the fact that they are traceless operators and that there are no initial correlation between them since the initial state is of a product form. So, the only non-vanishing contribution comes whenever the two-$\Gamma$s are in the same sub-system. Consequently, the $I$ matrix has a diagonal form such that
	\begin{eqnarray} 
	I=\left(\begin{array}{cccc}
	\ddots & 0 & &\\
	0 & I_{l-1} &0 &\\
	 & 0 &I_l &  0\\
	 & & 0&\ddots \\
	\end{array}\right),
	\label{eq328}
	\end{eqnarray}
where $I_l$ is a $L_l\times\ L_l$ matrix with $L_l$ the size of the $l$th subsystem, with matrix elements
\begin{equation}
(I_l)_{k,k'}=-\sum_{q=1}^{L_l}\psi^l_q(k)\phi^l_q(k')\tanh\left(\frac{\beta_l\epsilon^l_q}{2}\right)
	\label{eq329}
\end{equation}
where the $l$-labeled quantities refer to the $l$th subsystem Hamiltonian ${\cal H}_l$.
%%%%%%%%%%%%%%%%%%%%%%%%%%%%%%%%%%%%%%%%%%%%%%%%%%%%%%%%%%%%%%%%%%%%%%%%%%%%%%%%

%%%%%%%%%%%%%%%%%%%%%%%%%%%%%%%%%%%%%%%%%%%%%%%%%%%%%%%%%%%%%%%%%%%%%%%%%%%%%%%%
\section{System in contact with an infinite temperature ``bath"}
\subsection{Transverse magnetisation Green function}
We start our analyse by considering a chain where half of it (the left) is thermalized at inverse temperature $\beta_b$, and is  called the bath, and where the right half, the system, is initially at inverse temperature $\beta_s\equiv\beta$. Consequently, the initial matrix $\bf I$ is
	\begin{eqnarray} 
	\mathbf{I}=\left(\begin{array}{cc}
	{\bf I}_b& 0 \\ 
	0 & {\bf I}_s	 
	\end{array}\right)\; .
	\label{eq411}
	\end{eqnarray}
Taking the infinite bath temperature limit, the initial matrix ${\bf I}_b$ vanishes so that only the system part remains with 
\begin{equation}
(I_s)_{k,k'}=-\sum_{q=1}^{L_s}\psi^s_q(k)\phi^s_q(k')\tanh\left(\frac{\beta\epsilon^s_q}{2}\right)
\label{eq412}
\end{equation}
where $L_s$ is the system size. The system initial matrix can be decomposed into a surface contribution, near the interface boundary, and a homogeneous bulk part:
  	\begin{eqnarray}
	(\mathbf{I_s})_{k,k'}=(\mathbf{I_s}^{Bulk})_{k,k'}+(\mathbf{I_s}^{Surf})_{k,k'}
	\label{eq413}
	\end{eqnarray}
with
 	\begin{eqnarray}
	(\mathbf{I_s}^{Bulk})_{k,k'}=\sum_{n=1}^{L_s}\frac{(-1)^{k+k'}}{L+1}\cos\left(q_n(k-k')\right)\tanh\left(\beta\epsilon_{q_n}\right/2)\\
	(\mathbf{I_s}^{Surf})_{k,k'}=-\sum_{n=1}^{L_s}\frac{(-1)^{k+k'}}{L+1}\cos\left(q_n(k+k')\right)\tanh\left(\beta\epsilon_{q_n}\right/2)\; .
	\label{eq414}
	\end{eqnarray}
In the large size limit, $L_s\rightarrow\infty$, 
 	\begin{eqnarray}
	(\mathbf{I_s}^{Bulk})_{k,k'}=(-1)^{k+k'}\int_{0}^{\pi}\frac{d\theta}{\pi}\cos\left((k-k')\theta\right)\tanh\left(\beta\epsilon_{h}(\theta)\right/2)\; ,
	\label{eq415}
	\end{eqnarray}
and 
 	\begin{eqnarray}
	(\mathbf{I_s}^{Surf})_{k,k'}=-(-1)^{k+k'}\int_{0}^{\pi}\frac{d\theta}{\pi}\cos\left((k+k')\theta\right)\tanh\left(\beta\epsilon_{h}(\theta)\right/2)\; ,
	\label{eq416}
	\end{eqnarray}
for the surface contribution,
with $\epsilon_h(\theta)=h-\cos(\theta)$. 

\subsection{Zero temperature limit}
In the system low temperature limit, $T_s=0$, one has 
to distinguish between the region $h<1$ and $h\ge 1$ since $\tanh(\beta\epsilon_h(\theta)/2)={\rm sgn}(\epsilon_h(\theta))$.

\paragraph{\bf i) $h\ge1$ region}
For $h\ge 1$, the excitations are always positive and the sum in (\ref{eq412}) reduces to $-\sum_q\psi^s_q(k)\phi^s_q(k')=\delta_{k,k'}$, so that the initial matrix is just the identity matrix.  The average magnetisation, in the thermodynamic limit $L_{s,b}\rightarrow \infty$ is then given at later time by
	\begin{equation}
	m^z(n,t)\equiv \moyenne{\sigma_{n}^{z}}(t)=\sum_{k=-\infty}^{\infty} F^0_t(n-k) H(k)\; ,
	\label{eq421}
	\end{equation}
where $H$ is the Heaviside function. 
The Green function $G_t(n)=F^0_t(n)=J^2_{n}(t)$ can be expressed in the continuum limit as \cite{karevski1}  $G_t(n)\sim\frac{1}{t}g(n/t)$ with the scaling function
	\begin{equation}
	g(u)=\frac{1}{\pi\sqrt{1-u^2}} \ \ ,|u|\le 1
	\label{eq422}
	\end{equation}
and zero otherwise. Finally, with the step-like initial condition we have in the continuum limit
	\begin{eqnarray}
	m^z(x,t)=\left\{
	\begin{array}{cc}
	0 &\quad x<-t \nonumber\\
	\frac{1}{\pi}\arcsin(x/t)+\frac{1}{2}&\quad -t\le x\le t \\
	1 &\quad t<x.\nonumber
	\end{array}\right. \; .
	\label{eq423}
	\end{eqnarray}
This solves the extrem temperature set-up for $h\ge 1$.

\paragraph{\bf ii) $h<1$ region}
For $h<1$ one has a $k$-independent profile for the bulk contribution	
	\begin{eqnarray}
	p^{j}\equiv(\mathbf{I_s}^{Bulk})_{k,k+j}=(-1)^{j+1}\frac{2}{\pi}\frac{\sin(j\theta_F)}{j} \quad \forall j\ne 0\; ,
	\label{eq431}
	\end{eqnarray}
where $\theta_F=\arccos(h)$, and for the diagonal part, $j=0$, 
	\begin{eqnarray}
	p^{0}\equiv m^z(0)\equiv(\mathbf{I_s}^{Bulk})_{k,k}=\frac{2}{\pi}\arcsin(h)\;.
	\label{eq432}
	\end{eqnarray}
The surface term leads to 
	a $k$-decaying  part
 	\begin{eqnarray}
	(\mathbf{I_s}^{Surf})_{k,k+j}=(-1)^{j}\frac{2}{\pi}\frac{\sin((2k+j)\theta_F)}{2k+j}\quad  \forall j\; .
	\label{eq433}
	\end{eqnarray}
Neglecting the boundary effect term, $I^{Surf}$, and using the fact that $F_t^{2j+1}(n)=0$, one has for the transverse magnetisation
\begin{equation}
	\moyenne{\sigma_{n}^{z}}(t)=\left(\left[\sum_{j} \frac{p^{2j}}{m^z(0)} F_t^{2j}\right]\star m^z(0)H\right)(n) =\left(G_t\star m^z(0)\right)(n)\; ,
	\label{eq434}
\end{equation}
where we have used the fact that $(\mathbf{I}^{Bulk})_{k,k+2j}=P^{2j}(k)=p^{2j}H(k)$. The Green function $G_t$ is determined by the spatial derivative of the former equation, since 
$[m^z(x,t)]'=G_t\star m^z(0) H'=m^z(0) G_t\star \delta=m^z(0) G_t$. 
In the asymptotic limit (see the appendix) one obtains
	\begin{eqnarray}
	G_t(x)= \frac{1}{t}g\left(\frac{x}{t}\right)=\frac{1}{t}\frac{1}{2  \arcsin(h)\sqrt{1-(x/t)^2}}\;,
 	\quad \sqrt{1-h^2}\le |\frac{x}{t}|\le 1\; ,
	\label{eq435}
	\end{eqnarray}
and zero otherwize.\footnote{Note that the same analysis is handable in the Ising case at critical transverse field. The result is simpler with a Green function given by 
$
G_t(x)=\frac{1}{t}g^{Ising}\left(\frac{x}{t}\right)=\frac{1}{2t}\Pi\left(\frac{x}{2t}\right)
$
where $\Pi(u)=1 \;\forall u\in[-1/2,1/2]$ and zero otherwize. Details of the derivation are given in appendix~3.}
This disconnected support of the Green function will play an important role when considering the relaxation of a finite initially magnetised domain. It will lead to the appearance of a finite relaxation time.

\begin{figure}
\epsfxsize=9cm
\begin{center}
\mbox{\epsfbox{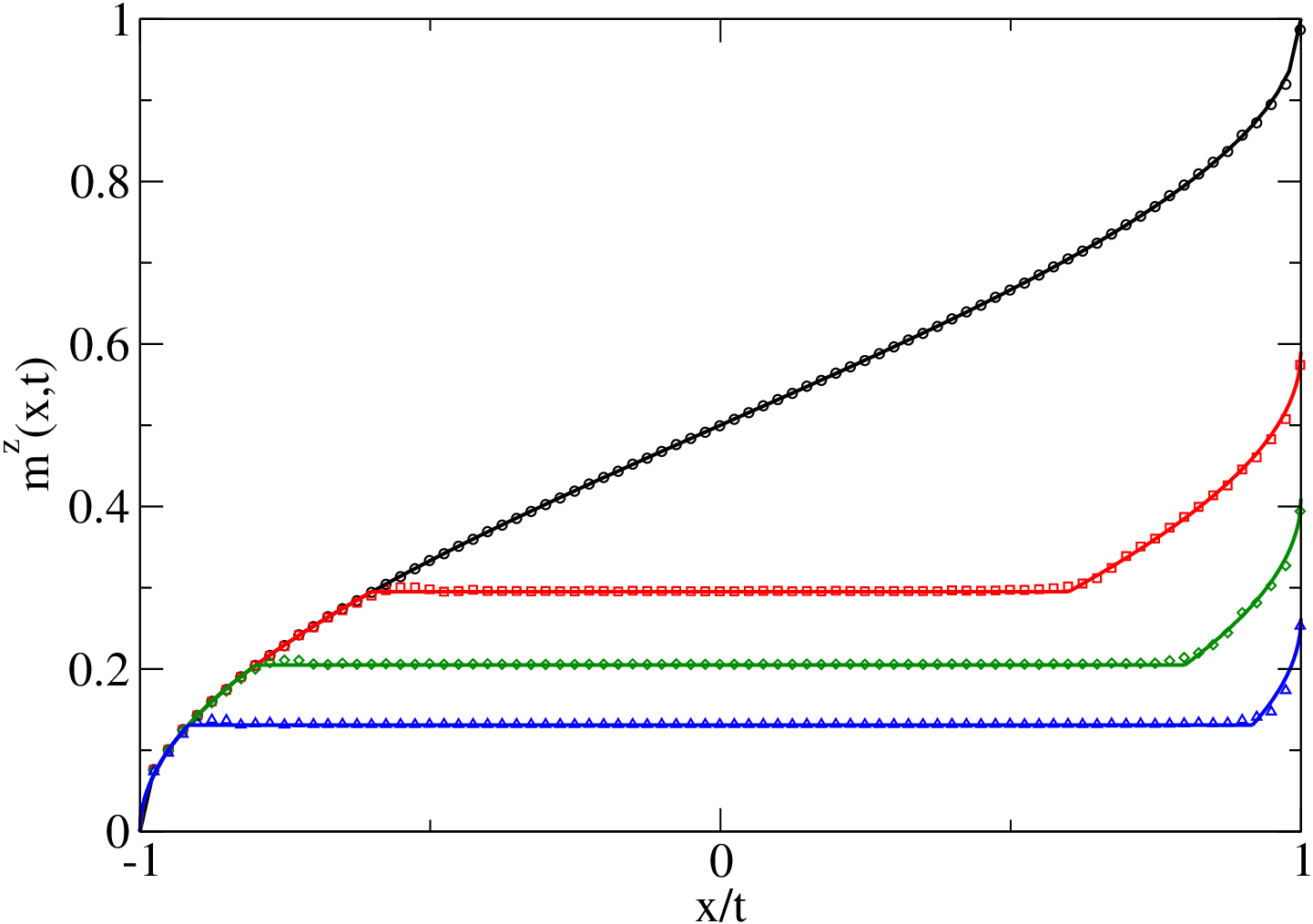}}
\end{center}
\caption{Magnetization profiles with $T_s=0$, $T_b=\infty$ for different fields $h=1$, $h=0.8$, $h=0.6$ and $h=0.4$ from top to bottom on a chain of total size $L=500$. The analytical results are in full lines.
\label{Front1}
}
\end{figure}

It turns out that the step-like initial state generated by the two-temperature set-up develops a plateau of constant magnetisation,  as it was already observed in (\cite{antal3,ogat}), due to the vanishing of the Green function in the domain $|x/t|<\sqrt{1-h^2}$: 
\begin{eqnarray}
	m^z(x,t)=\Phi_h(x/t)+m^z(0)/2
\label{eq436}
\end{eqnarray}
with $\Phi_h(-u)=-\Phi_h(u)$ where
	\begin{eqnarray}
	\Phi_h(u)=\left\{
	\begin{array}{cc}
	0 &\quad 0<u\le \sqrt{1-h^2}\nonumber\\
	m^z(0)/2-\arccos(u)/\pi &\quad \sqrt{1-h^2}<u\le 1 \\
	m^z(0)/2 &\quad 1<u\nonumber
	\end{array}\right. \; .
\label{eq437}
	\end{eqnarray}	
We show the numerical and analytical results in figure (\ref{Front1}) for different fields. 
 %%%%%%%%%%%%%%%%%%%%%%%%%%%%%%%%%%%%%%%%%%%%%%%%%%%%
%%%%%%%%%%%%%%%%%%%%%%%%%%%%%%%%%%%%%%%%%%%%%%%%%%		
\subsection{Finite temperature effects}
\begin{figure}
\epsfxsize=9cm
\begin{center}
\mbox{\epsfbox{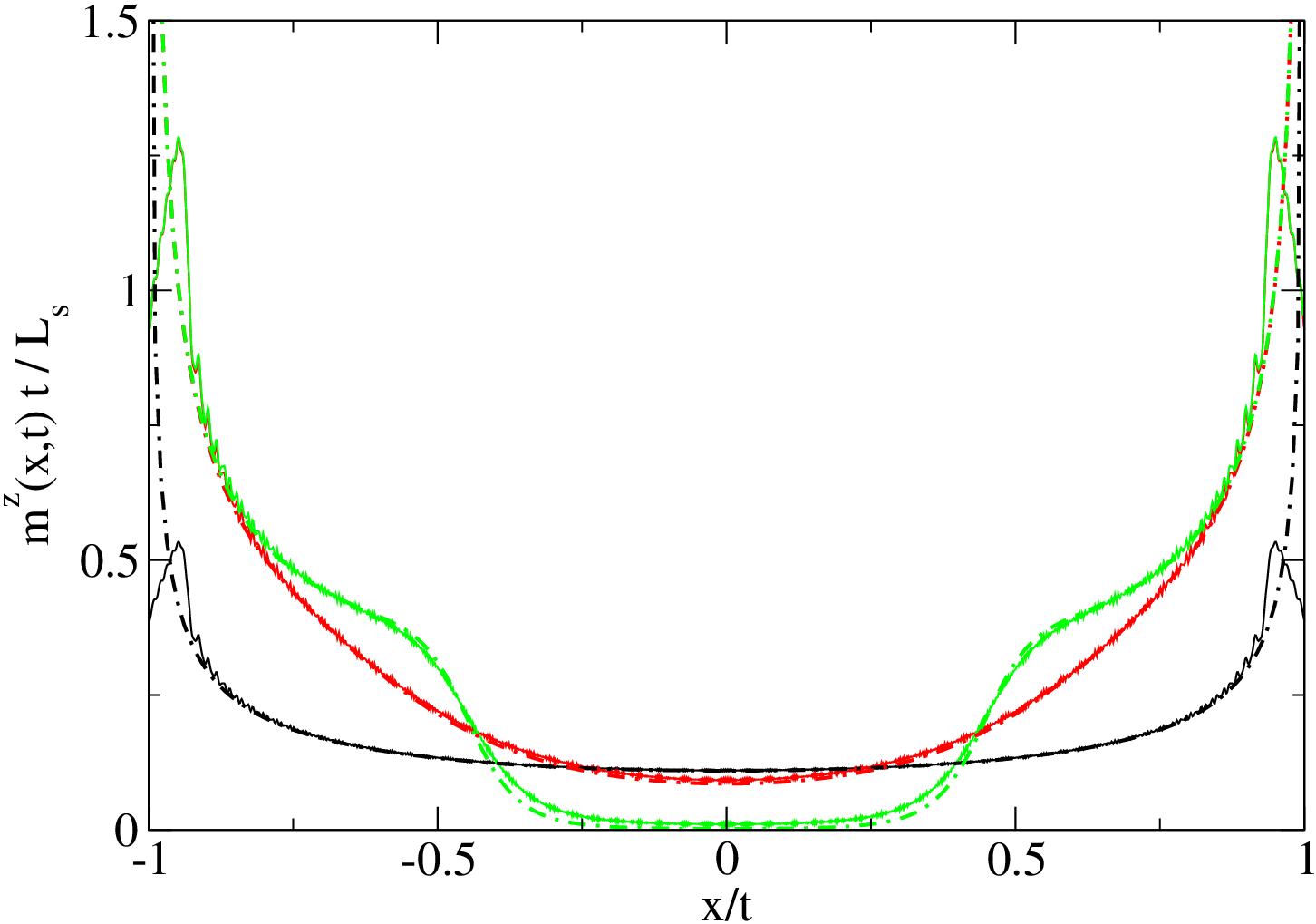}}
\end{center}
\caption{Numerical rescaled magnetisation profiles for three different inverse temperatures (At $x/t=0$ from top to bottom $\beta_s=1$, $\beta_s=10$, $\beta_s=100$) at $h=0.8$. The size of the system is $L_s=100$ and the time is $t=700$. The dashed lines are the corresponding analytical curves for the Green functions.
\label{GreenT}
}
\end{figure}

While keeping a very hot bath (formally taken to be at infinite temperature) we now consider the effect of a finite initial temperature system state on the relaxation of the transverse magnetisation. As in the previous section, we can compute the Green function in the asymptotic limit from the step-like initial condition. One finds for $h\ge 1$ (see the appendix for details) after a lengthy derivation
\begin{eqnarray}
G_t(x)=\frac{1}{2\pi t  m^z(0)\sqrt{1-\left(\frac{x}{t}\right)^2}}&\left\{
\tanh\frac{\beta_s}{2}\left(h+\sqrt{1-\left(\frac{x}{t}\right)^2}\right)\right.\nonumber\\
&\qquad \left.+\tanh\frac{\beta_s}{2}\left(h-\sqrt{1-\left(\frac{x}{t}\right)^2}\right)\right\}
\label{eq441}
\end{eqnarray}
where $\beta_s$ is the inverse initial system temperature and where
\begin{equation}
m^z(0)=\frac{1}{\pi}\int_0^\pi {\rm d}\theta \tanh\left(\frac{\beta_s}{2}(h-\cos\theta)\right)
\label{eq442}
\end{equation}
is the initial bulk magnetisation density. Expression~(\ref{eq441}) is also valid at $h<1$. 
On figure~(\ref{GreenT}) we have plotted the numerical magnetisation profiles compared with analytical result~(\ref{eq441}) for different temperatures, since in the long time limit $t\gg L_s$ one recovers the Green function as 
$m^z(x,t)=G_t(x)\star\frac{m^z(0)L_s}{t}\delta(x/t)$. 
%%%%%%%%%%%%%%%%%%%%%%%%%%%%%%%%%%%%%%%%%%%%%%%%%%%%
%%%%%%%%%%%%%%%%%%%%%%%%%%%%%%%%%%%%%%%%%%%%%%%%%%		

One may notice that it is also possible to deduce the Green function in a very simple way starting with the kink-like initial state and assuming that the magnetisation at the interface takes for all times its asymptotic value $m^z(0)/2$, since the initial bath magnetisation is zero while the initial system magnetisation is $m^z(0)$. Indeed, using (\ref{eq434}), which is in fact also valid at $h\ge 1$, one has 
$m^z(x,t)=(G_t\star m^z(0) H)(x)$.  Together with the scaling assumption $G_t(x)=\frac{1}{t m^z(0)}g(x/t)$ one has
\begin{equation}
m^z(x,t)=\int_{-1}^xdu\;g(u)\; .
\end{equation}
Now, taking into account the fact that at the interface $x=0$, the magnetisation is given by (\ref{eq442}), one has
\begin{equation}
m^z(0,t)=\int_{-1}^0 du\;g(u)=\frac{m^z(0)}{2}=\int_0^{\pi}\frac{d\theta}{2\pi}\tanh\left(\frac{\beta_s}{2}(h-\cos\theta)\right)\; .
\end{equation}
Decomposing the integral into two peaces, one from $0$ to $\pi/2$ and the other one from $\pi/2$ to $\pi$, by a simple change of the integration variable one obtains
\begin{equation}
\int_{-1}^0 du\; g(u)=\int_{-1}^0 du
\frac{\tanh\left(\frac{\beta_s}{2}(h-\sqrt{1-u^2})\right)+\tanh\left(\frac{\beta_s}{2}(h+\sqrt{1-u^2})\right)}
{2\pi\sqrt{1-u^2}}
\end{equation}
which gives by identification $g(u)$ and so $G_t(x)$.

\section{Relaxation of an initial droplet}
If one considers initially a finite system of size $L_s$ (the initial droplet) at an inverse temperature $\beta_s$ in contact at both ends with infinite temperature ``baths'', from the convolution of the asymptotic Green function  with $P^0(k)=m^z(0)\Pi\left(\frac{k}{L_s}\right)$ one obtains the transverse magnetisation at later times.
The total droplet magnetisation density $m^z(t)=\frac{1}{L_s}\sum_{k\in L_s} m^z(k,t)$ will relaxe toward zero since the baths are at infinite temperature (and consequently have a vanishing transverse magnetisation). From the asymptotic Green function obtained in the different regimes $h\ge1$ and $h<1$ we obtain two different relaxation behaviours.

\subsection{Zero temperature droplet}
In the $h\ge 1$ regime the droplet magnetisation density behaves as 
\begin{eqnarray}
m^z(t)=\left\{\begin{array}{l}
1-\frac{2}{\pi}\frac{t}{L_s}\quad t\le \tau_1\\
\frac{2}{\pi}\arcsin\left(\frac{L_s}{t}\right)-\frac{2}{\pi}\frac{t}{L_s}\left(1-\sqrt{1-\left(\frac{L_s}{t}\right)^2}\right)\quad t\ge \tau_1
\end{array}
\right.\; ,
\label{eq511}
\end{eqnarray}
with $\tau_1=L_s$.
Asymptotically the magnetisation relaxes with a power law behaviour \cite{schutztrimper}
\begin{equation}
m^z(t)\simeq \frac{L_s}{\pi t}\quad h\ge 1\; .
\label{eq512}
\end{equation}
One may notice that in the critical Ising chain one has exactly the same relaxation behaviour for the droplet magnetization: $m^z(t)\simeq \frac{L_s}{\pi t}$ \cite{platini1,platini2}.

On the contrary, in the low field regime, $h<1$, the time behaviour of the droplet magnetisation density obtained from the asymptotic Green function~(\ref{eq435}) is given by
\begin{eqnarray}
m^z(t)&=\frac{2}{\pi}\arccos\left(\sqrt{1-h^2}\right)-\frac{2}{\pi}\frac{t}{L_s}h,\quad t\le \tau_1\nonumber\\
m^z(t)&=\frac{2}{\pi}\arccos\left(\sqrt{1-h^2}\right)-\frac{2}{\pi}\arccos \left(\frac{L_s}{t}\right) \nonumber\\
&\qquad-\frac{2}{\pi}\frac{t}{L_s}\left(h-\sqrt{1-\left(\frac{L_s}{t}\right)^2}\right),
\quad \tau_1\le t\le \tau
\label{eq513}
\end{eqnarray}
and
\begin{equation}
m^z(t)=0,\quad t\ge \tau_2\; .
\label{eq514}
\end{equation}
with $\tau_2=L_s/\sqrt{1-h^2}=L_s/\cos(m^z(0)\pi/2)$.
The vanishing of the magnetisation at times $t\ge \tau_2$ is a direct consequence of the vanishing of the Green function in the region $-\sqrt{1-h^2}<x/t<\sqrt{1-h^2}$. However, this exact cancellation is valid only in the limit $L_s\rightarrow \infty$. 
For a finite droplet, one has finite size corrections to the Green function such that at the leading order the Green function is given by
$G_t(x)=g(x/t)/t[1+{\cal O}(1/L_s)]$. These corrections lead for the magnetisation density to a size independant correction to zero of order $1/t$. Consequently, the ratio between the magnetization densities at $h< 1$ and $h\ge 1$ is asymptotically (in time) given by
\begin{equation}
\frac{m^z_{h< 1}(t)}{m^z_{h\ge 1}(t)}={\cal O}\left( \frac{1}{L_s} \right)\; .
\end{equation}
It is in this sense that the magnetisation density within the droplet changes from a slow relaxation (\ref{eq512}), depending on the size of the system, to a much faster relaxation since it is independent of the size. If one considers the total remaining magnetisation of the droplet, we pass from a diffusive like behaviour $L_s^2/t$ at $h\ge 1$ to a balistic-like behaviour $L_s/t$ at $h<1$. These two regimes reflect the fact that the dispersion relation changes from a parabolic shape, $\epsilon(k)\sim h-1 + k^2/2$ at $h\ge 1$ to a linear regime $\epsilon(k)\sim |k-k_F|$ for $h<1$. 

\begin{figure}
\epsfxsize=9cm
\begin{center}
\mbox{\epsfbox{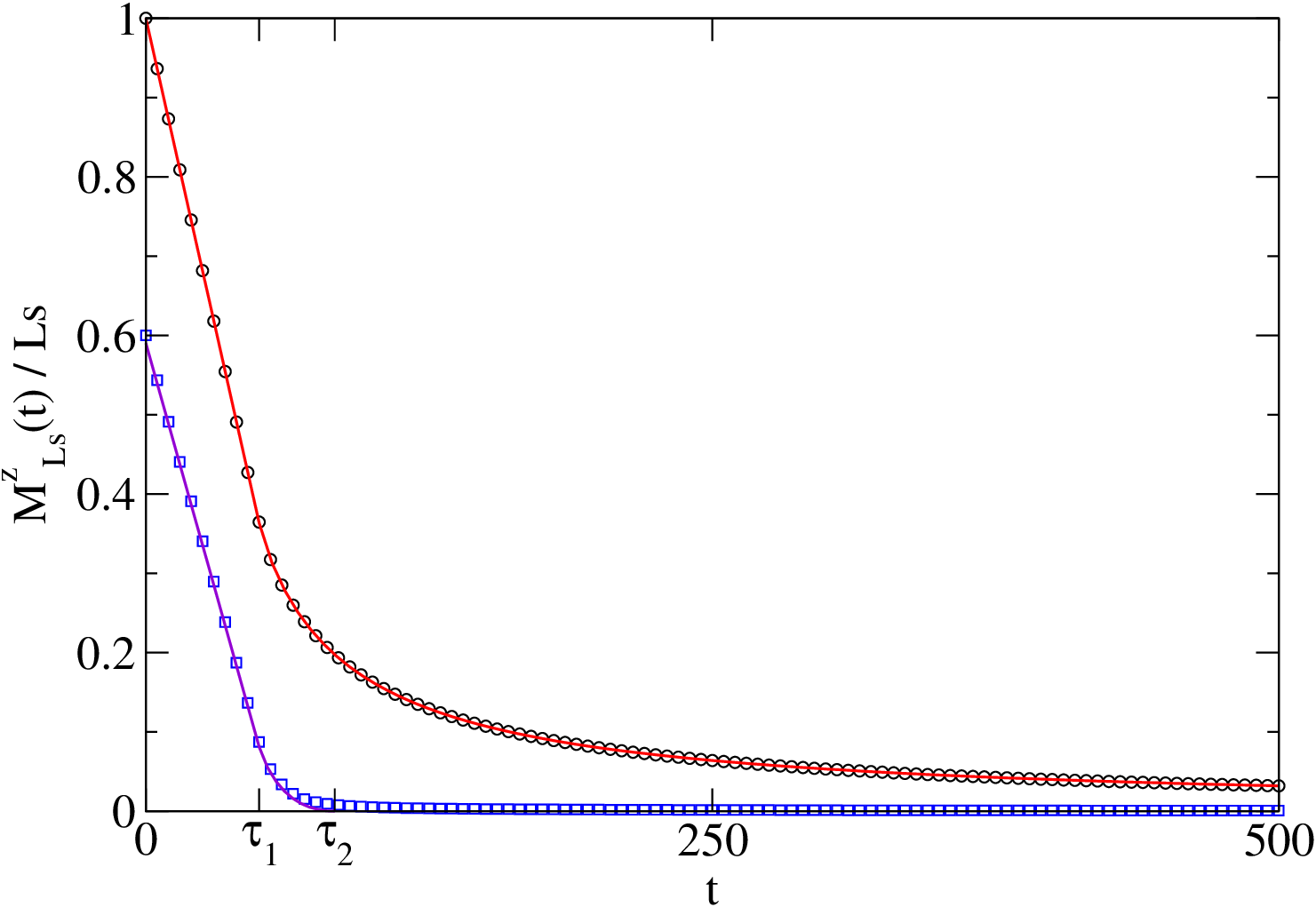}}
\end{center}
\caption{Relaxation of the magnetisation density at fields, from top to bottom, $h=1$ and $h=0.8$ for an initial droplet of size $L_s=50$. The full lines correspond to the analytical results. 
\label{Mz-temps}
}
\end{figure}

On figure (\ref{Mz-temps}) we show the numerical results compared to the analytical expressions obtained for the droplet magnetisation density. The agreement between the exact numerical results and the analytical asymptotic expressions is excellent.
 %%%%%%%%%%%%%%%%%%%%%%%%%%%%%%%%%%%%%%%%%%%%%%%%%%%%
%%%%%%%%%%%%%%%%%%%%%%%%%%%%%%%%%%%%%%%%%%%%%%%%%%		
	
\subsection{Finite temperature droplet}
With the help of the finite temperature Green function~(\ref{eq441}), one may compute the droplet magnetisation density at the leading order:
\begin{equation}
m^z(t)\simeq \frac{L_s}{\pi t}\frac{1}{2} \left(\tanh\left[\frac{\beta_s}{2}(h+1)\right]+
\tanh\left[\frac{\beta_s}{2}(h-1)\right]\right)
\label{eq521}
\end{equation}
which shows the $L_s/t$ power law decay for any transverse field $h$. In the $h<1$ region, the faster relaxation is recovered in the low temperature limit $\beta_s\rightarrow\infty$.  Indeed, for sufficiently small temperature the magnetisation density is given by
\begin{equation}
m^z(t)\simeq \frac{L_s}{\pi t}\exp\left(-\beta_s(1-h)\right)
\label{eq522}
\end{equation}
and one may identify a typical size-dependent inverse temperature $\beta(L_s)\equiv \ln L_s/(1-h)$ and one recovers the fast relaxation for inverse temperatures $\beta_s\gg \beta(L_s)$.

\section{Summary and discussion}
%%%%%%%%%%%%%%%%%%%%%%%%%%%%%%%%%%%%%%%%%%%%%%%%%%%%%%%%%%%%%%%%%
We have presented here analytical results, confirmed by exact numerical calculations, of the relaxation of the transverse magnetisation of the XX quantum chain. In particular, starting from an initial product state depending on the strength of the initial magnetic field, we have obtained the asymptotic Green function associated to the transverse magnetisation. When the system is divided initially into two parts, a hot bath and a cold sub-system, we have shown that the Green function takes the scaling form $G_t(x)=\frac{1}{t}g(x/t)$, where the explicit shape of the scaling function $g(u)$ depends on the initial temperatures and on the field. In the extreme scenario, with a vanishing sub-system temperature and an infinite bath temperature, at low fields, $h<1$, the scaling function shows the very peculiarity that its support is disconnected. This property leads to the emergence of a flat magnetization profile when one starts with a kink-like initial profile \cite{antal3}. In the case of an initial finite magnetised domain it leads to the fact that its relaxation is much faster in the low field regime, $h<1$ than it is at higher fields. Namely, one has $m_{h<1}(t)/m_{h\ge1}(t)\sim 1/L_s$ where $m(t)$ is the domain magnetisation density at time $t$ and  where $L_s$ is the size of the domain. 

Finally, one may notice that for a Green function taking the scaling form $G_t(x)=\frac{1}{t}g(x/t)$, one has identically
\begin{equation}
\partial_t G+\partial_x \frac{x}{t} G=0
\label{eq61}
\end{equation}
so that the impulsional response magnetisation $m^z_i=m^z(0)G_t$ decays in time and satisfy the continuity equation
\begin{equation}
\partial_t m^z_i(x,t)+\partial_x j_i(x,t)=0
\label{eq62}
\end{equation}
with the current $j_i$ simply given by $j_i(x,t)=\frac{x}{t}m^z_i(x,t)$. 
In the step-like situation, the scaling form $\frac{1}{t}g(x/t)$ of the Green function leads to a scaling solution $m^z_s(x,t)=\Phi(x/t)$ which obviously satisfy the homogeneous scaling equation
\begin{equation}
(t\partial_t+x\partial_x)m^z_s(x,t)=0
\label{eq63}
\end{equation}
which leads for the current $j_s$ to the equation $\partial_x j_s=\frac{x}{t}\partial_x m^z_s$. So that the step-like current is related to the impulsional current through
$
\partial_x j^s=j^i
$
since $\partial_x m^z_s=m^z_i$.

\section*{Acknowledgements}
We want to thank the Groupe de Physique Statistique of the Laboratoire de Physique des Mat\'eriaux for usefull and pleasant discussions. Gunter Sch\"utz is gratefully acknowledged for his hospitality and support during this summer in the Forschungszentrum J\"ulich where part of this work was done.

\section*{Appendix 1: Dynamics of the lattice operators }
The Clifford operators $\Gamma_n^1$, $\Gamma_n^2$ can be expressed in terms of fermionic operators.
	\begin{eqnarray}
	\Gamma_n^1(t)&=&\sum_{q=1}^{L}\phi_q(n)(\eta_q^{\dag}(t)+\eta_q(t)) \nonumber \\
	\Gamma_n^2(t)&=&\sum_{q=1}^{L}i\psi_q(n)(\eta_q^{\dag}(t)-\eta_q(t))
	\end{eqnarray}
The time evoution of the creation and annihilation operators is obtained by $\eta_q^{\dag}(t)=U_q^{\dag}(t)\eta_q^{\dag}U_q(t)$ with
	\begin{eqnarray}
	U_q(t)=e^{-i\epsilon_q\eta_q^{\dag}\eta_qt} 
	\end{eqnarray}
and leads to
	\begin{eqnarray}
	\eta_q^{\dag}(t)&=&e^{i\epsilon_qt}\eta_q^{\dag}\nonumber \\
	\eta_q(t)&=&e^{-i\epsilon_qt}\eta_q \; .
	\end{eqnarray}	
With the help of these relations we can re-expressed the Cliffords operators in terms of the initial time operator $\Gamma$. Finally one obtains
	\begin{eqnarray} 
	\Gamma_n^j(t)=\sum_{k,i} \langle\Gamma_{k}^{i} |\Gamma_{n}^{j}\rangle_t \Gamma_k^i
	\end{eqnarray}
with
	\begin{eqnarray} 
	\langle\Gamma_{k}^{1} |\Gamma_{n}^{1}\rangle_t&=&\sum_q\phi_q(k)\phi_q(n)\cos(\epsilon_qt)\nonumber \\
	\langle\Gamma_{k}^{2} |\Gamma_{n}^{2}\rangle_t&=&\sum_q\psi_q(k)\psi_q(n)\cos(\epsilon_qt)\nonumber \\
	\langle\Gamma_{k}^{1} |\Gamma_{n}^{2}\rangle_t&=&\langle\Gamma_{n}^{2} |\Gamma_{k}^{1}\rangle_{-t}=-\sum_q\phi_q(k)\psi_q(n)\sin(\epsilon_qt)\; .
	\end{eqnarray}
	
\section*{Appendix 2: Zero temperature XX Green function}
The contribution of the bulk to the magnetisation is given by:
  	\begin{eqnarray}
	m^z(n,t)=\sum_{j}\sum_{k}F^{j}_{t}(n-k)(\mathbf{I_s}^{Bulk})_{k,k+j}\; .
	\end{eqnarray}
Using  $(\mathbf{I_s}^{Bulk})_{k,k+j}=H(k)p^{j}$ and the relation $(\mathbf{I_s}^{Bulk})_{k,k+j}=(\mathbf{I_s}^{Bulk})_{k+j,k}$ one has:
 	\begin{eqnarray}
	m^{z}(n,t)=m^z(0)\sum_{k=0}^{\infty}J_{n-k}^2(t)+2\sum_{k=0,j=1}^{\infty}(-1)^jJ_{n-k}(t)J_{n-k-2j}(t)p^{2j}\; .
	\end{eqnarray}
with $p^0=m^z(0)$. 
The spatial derivative $\Phi'(n+1,t)=m^{z}(n+1,t)-m^{z}(n,t)$ is 
 	\begin{eqnarray}
	\Phi'(n,t)=J_{n}(t)\left(m^z(0)J_{n}(t)+2\sum_{j=1}^{\infty}(-1)^jp^{2j}J_{n-2j}(t)\right)
	\end{eqnarray}
Using the asymptotic expansion of the Bessel function for $\nu>>1$
	\begin{eqnarray}
	J_{n}(t)=J_\nu\left(\frac{\nu}{\cos\beta}\right)=\sqrt{\frac{2}{\pi\nu\tan\beta}}\cos\psi\; ,
	\end{eqnarray}
with $\psi=\nu(\tan\beta-\beta)-\pi/4$, and taking $n\rightarrow\infty$  one has  
	\begin{eqnarray}
	\Phi'(n,t)&\simeq&\frac{2}{\pi t\sqrt{1-(\frac{n}{t})^2}}\times\nonumber\\
&&\left\{\cos^2(\psi)\left[m^z(0)+2\sum_{j\ge1}(-1)^jp^{2j}\cos(2j\arccos(\frac{n}{t}))
\right]\right.\nonumber\\
&&\left.\qquad-\cos(\psi)\sin(\psi)\left[2\sum_{j\ge1}(-1)^jp^{2j}
\sin(2j\arccos(\frac{n}{t}))\right]\right\}\; .
	\end{eqnarray}
Keeping only the first term in this expression, where $\cos^2(\psi)$ is replaced by its average value $1/2$ and taking the explicit expressions for the profiles 
$p^j$,  given in (\ref{eq431}), together with 
 	\begin{eqnarray}
	\sum_{j=1}^{\infty}(-1)^j\frac{\sin(jx)}{j}=-x/2& -\pi<x<\pi\; ,
	\end{eqnarray}
 	\begin{eqnarray}
	\sum_{j=1}^{\infty}(-1)^j\frac{\sin(jx)}{j}=\pi-x/2& \pi<x<3\pi\; ,
	\end{eqnarray}
one obtains finally 
	\begin{eqnarray}
	\Phi'(n,t)=\frac{1}{\pi t\sqrt{1-\nu^2}}\left\{
	\begin{array}{cc}
	0 &\quad n/t\le \sqrt{1-h^2}\nonumber\\
	1 &\quad \sqrt{1-h^2}<n/t\le 1.
	\end{array}\right.
	\end{eqnarray}	
for positive $n$ and $\Phi'(-n,t)=\Phi'(n,t)$. The Green function is simply given by $G(x,t)=\Phi'(x,t)/m^z(0)$. 

\section*{Appendix 3: Critical Ising Green function}	
In the Ising case the bulk contribution is given by (see \cite{platini1,platini2}):
  	\begin{eqnarray}
	m^z(n,t)=\sum_{j}\sum_{k}P^{k,k+j}_{n}(t)(\mathbf{I}^{Bulk}_s)_{k,k+j},
	\end{eqnarray}
with   
	\begin{eqnarray}
	(-1)^jP^{k,k+j}_{n}(t)&=&J_{2(n-k)}(2t)J_{2(n-k)-2j}(2t)\nonumber\\
	&&\qquad-J_{2(n-k)+1}(2t)J_{2(n-k)-2j-1}(2t)\; .
	\end{eqnarray}
At system vanishing temperature $T_s=0$ the $I_s$ elements are given by:
	\begin{eqnarray}
	(\mathbf{I_s}^{Bulk})_{k,k'}=\frac{2}{\pi}\frac{(-1)^{k'-k}}{2(k'-k)+1}\; .
	\end{eqnarray}
Using the relation $(\mathbf{I_s}^{Bulk})_{k,k'}=(\mathbf{I_s}^{Bulk})_{k'+1,k}$ one has:
 	\begin{eqnarray}
	m^z(n,t)=\sum_{j=0}^{\infty}\sum_{k}F^{j}_t(n-k)(\mathbf{I}^B)_{k,k+j}
	\end{eqnarray}
with 
	\begin{eqnarray}
	(-1)^jF^{j}_t(p)&=&2\left(J_{2p}'(2t)J_{2p-2j-1}(2t)
	-J_{2p}(2t)J_{2p-2j-1}'(2t)\right) \; .
	\end{eqnarray}
The spatial derivative function $\Phi'(n+1,t)=m^{z}(n+1,t)-m^{z}(n,t)$ is then given by
	\begin{eqnarray}
	\Phi'(n,t)=\frac{4}{\pi}J_{2n}'(2t)\sum_{j=0}^{\infty}\frac{J_{2n-2j-1}(2t)}{2j+1}-\frac{4}{\pi}J_{2n}(2t)\sum_{j=0}^{\infty}\frac{J_{2n-2j-1}'(2t)}{2j+1}\; .
	\end{eqnarray}	
Taking this expression together with the asymptotic expansion of the Bessel function we finally arrive at 
	\begin{eqnarray}
	G(x,t)=\frac{1}{2t}\Pi\left(\frac{x}{2t}\right).
	\end{eqnarray}		

\section*{Appendix 4: Finite temperature XX Green function}
In the system non zero temperature case, one has
 	\begin{eqnarray}
	\Phi'(n,t)=m^z(0)J_{n}^2(t)+J_{n}(t)A_n(t)
	\end{eqnarray}
with
 	\begin{eqnarray}
	A_n(t)=2\sum_{j=1}^{\infty}(-1)^jp^{2j}J_{n-2j}(t)
	\end{eqnarray}
where $p^{2j}H(k)=(I^{Bulk})_{k,k+2j}$ is defined in (\ref{eq415}). 
Using the relation
	\begin{eqnarray}
	 \tanh(x)=1+2\sum_{k=1}^{\infty}(-1)^ke^{-2kx}\qquad \forall x\ge 0
	\end{eqnarray}	
(imposing $h\ge1$) one has:
	\begin{eqnarray}
	 p^{2j}=I_{2j}(0)+2\sum_{k=1}^{\infty}(-1)^ke^{-2\beta_s hk}I_{2j}(\beta_sk)&
	\end{eqnarray}	
where $I_{2j}(u)$ are the modified Bessel functions. Then 
 	\begin{eqnarray}
	A_n(t)=2\sum_{j=1}^{\infty}(-1)^j2\sum_{k=1}^{\infty}(-1)^ke^{-2\beta_s hk}I_{2j}(\beta_sk)J_{n-2j}(t)
	\end{eqnarray}
The asymptotic behaviour of the Bessel function leads to:
 	\begin{eqnarray}
	J_{n}(t)A_n(t)&=&\frac{2}{\pi t\sqrt{1-(n/t)^2}}\sum_{k=1}^{\infty}(-1)^ke^{-2\beta_s hk}\times\nonumber\\
	&&\qquad 2\sum_{j=1}^{\infty}(-1)^jI_{2j}(\beta_sk)\cos(2j\arccos(n/t)).
	\end{eqnarray}
Since $I_{2j}(u)=(-1)^jJ_{2j}(iu)$ one has
 	\begin{eqnarray}
	2\sum_{j=1}^{\infty}(-1)^jI_{2j}(\beta_sk)&\cos\left(2j\arccos(n/t)\right)=2\sum_{j=1}^{\infty}J_{2j}(i\beta_sk)\cos\left(2j\arccos(n/t)\right)\nonumber\\
	&=\cosh\left(\beta_s k\sqrt{1-(n/t)^2}\right)-J_0(i\beta_s k)
	\end{eqnarray}
Finally, reinjecting this expression into the previous one, one obtains
 	\begin{eqnarray}
	\Phi'(x,t)=\frac{1}{2\pi t \sqrt{1-\left(\frac{x}{t}\right)^2}}
&\left\{
	\tanh\left(\frac{\beta_s}{2}\left[h+\sqrt{1-\left(\frac{x}{t}\right)^2}\right]\right)\right.\nonumber\\
	&\quad\left.+\tanh\left(\frac{\beta_s}{2}\left[h-\sqrt{1-\left(\frac{x}{t}\right)^2}\right]\right)\right\}\; .
	\end{eqnarray}

\end{document}